\documentclass[aps, preprint, showpacs, epsfig]{revtex4}
\usepackage{graphics}
\usepackage{graphicx}
\usepackage{epsfig}
\def \be {\begin{equation}}
\def \ee {\end{equation}}
\def \ba {\begin{eqnarray}}
\def \ea {\end{eqnarray}}
\def \bm {\begin{displaymath}}
\def \em {\end{displaymath}}
\begin{document}
\title{Dynamics of a bubble formed in double stranded DNA }
\author{Shikha Srivastava and Yashwant Singh}
\affiliation{Department of Physics, Banaras Hindu University, 
Varanasi-221 005,
India}
\date{\today}
\begin{abstract}
We study the fluctuational dynamics of a  tagged base-pair 
in double stranded DNA. We calculate the drift force which acts on the tagged
base-pair using a potential model that describes interactions at base pairs
level and use it to construct a Fokker-Planck equation.The calculated
displacement autocorrelation function is found to be in very good agreement 
with the experimental result of Altan-Bonnet {\it et. al.} Phys. Rev. Lett. {\bf 90},
138101 (2003) over the entire time range of measurement. We calculate
the most probable displacements which predominately contribute to the 
 autocorrelation function and the half-time history of these displacements.
 
\end{abstract}
\maketitle

DNA double stranded helical structure is stabilized by the hydrozen bonding 
between complementary bases and the stacking between neighbouring bases [1]. 
In physiological solvent conditions the average value of these interactions 
for a base pair is of the order of few $k_{B}T$ (thermal energy) [2] and thermal 
fluctuations can lead to local and transitory unzipping of the double 
strands [3,4]. The co-operative opening of a sequence of consecutive base pairs 
leads to formation of local denaturation zones (bubbles). As an AT base pair 
connected by two hydrogen bonds needs less energy to unzip compared to a GC 
base pair which is connected by three hydrogen bonds, initiation of a bubble
generally takes place in an AT rich region. A DNA bubble 
consists of flexible single stranded DNA and its size fluctuates by zipping
and unzipping of base pairs at the two zipper forks where the bubble connects
to the intact double strands. The average size of a bubble depends on the
sequence of base pairs, temperature and ionic strength and varies from few 
broken base pairs at room temperature to few hundred open base pairs close to
melting temperature [5,6].

The formations of bubble at room or physiological temperatures are rare and
intermittent with life times of the order of millisecond [4]. The occurrence 
of such bubble domains is important as the opening of dsDNA base pairs by 
breaking the hydrogen bonds between complementary bases disrupts the helical 
stack and may initiate biological processes of transcription, replication 
and protein binding [7,8]. From physics point of view, DNA bubbles offer a quasi
one -dimensional system for the study of fluctuational dynamics.

In an experiment by Altan-Bonnet {\it et. al.} [4] the dynamics of a single bubble 
in three synthetic DNA constructs having the same GC rich region but 
different AT base pairs regions have been investigated by fluorescence
correlation spectroscopy (FCS). In the middle of the AT region a T base pair 
was tagged with a fluorophore while the neighbouring T base of the other
strand was tagged with a quencher. The correlation spectrum of fluctuating base 
pairs was monitored. The dynamics was found to follow a multi-state relaxation
kinetics in a wide temperature range with a characteristic time scale
in the range of  $20-100\mu{s}$. Several theoretical
models [4,9,12,13]have recently been proposed to explain the observed multi-state 
breathing dynamics. In one of these models [9] the bubble free energy 
that corresponds to a bubble of infinitely large size [10,11] and which 
accuracy for a bubble of few broken base pairs, to best of our knowledge,
is not established has been used. Other theoretical models include discrete 
master equations and stochastic Gillespie schemes [4,12,13].

In this Letter, we develop a general theory to study the fluctuational
dynamics of a tagged base pair by means of a Fokker-Planck equation
based on a potential field which acts on the base pair and which we obtain by
integrating out the degrees of freedom of all base pairs of a dsDNA except 
those associated with the tagged one. We use the simple potential model 
of Peyard-Bishop-Dauxious (PBD) [14] to represent the interactions in dsDNA
at base pairs level.

The PBD model reduces the degrees of freedom of DNA to a one -dimensional
chain of effective atom compounds describing the relative base pair 
separation $y_i$ from the ground state position $y_i$=0. The potential
of the model is written as
\ba
U(y^{N})&=&\sum_i[D_{i}(e^{-2a_{i}y_{i}}-2e^{-a_{i}y_{i}}) \nonumber \\
&&+\kappa/2(1+\rho e^{-\alpha({y_i+y_{i-1}})}
{(y_i-y_{i-1}})^2)]
\ea

where N is the number of base pairs, summation on the r.h.s. is over all 
base pairs of the molecule and $y^N$=$\lbrace y_i \rbrace$, the set of relative base pair
separations. The first term of Eq.(1) is the Morse potential that represents
the hydrogen bonds between the bases of the opposite strands and the second 
term represents the stacking interaction between adjacent base pairs. The 
values of parameters found by Campa and Giansanti [15] are 
$\kappa=0.025eV\rm\AA^{-2}$,$\rho=2$ and $\alpha=0.35\rm\AA^{-1}$
for the stacking part, while for the Morse potential 
 $D_{AT}$ = 0.05 eV, $a_{AT}$ = $4.2\rm\AA^{-1}$ for an AT base pair and 
$D_{GC}$ = 0.075 eV and $a_{GC}$ = $6.9\rm\AA^{-1}$ for a GC base pair. 

We now consider one of the DNA molecules (named A18) investigated
by Altan-Bonnet {\it et. al.} [4] and take the $17{\it th}$ base pair
counted from the $5^\prime-$ end as the tagged base pair. The 
interactions in the molecule is represented by the PBD model.
We add a harmonic 
potential $u_{h}(y_N)$=$\theta(y{_N}-2)cy_{N}^2$ where 
$c\sim 1.0\times 10^{-2}\rm\AA^{-2}$ and $\theta(y)$ a Heaviside step
function at the terminal GC base pair to avoid the complete separation
of the two strands. In experiment [4] this was achieved by attaching 
a hairpin loop of 4T.  
The potential felt by the tagged base pair at a separation $y$ from 
the ground state $y=0$ is found from the relation
\ba
V(y)= -k_BT[lnZ_n(y_0)-lnZ_n(0)] 
\ea
where
\hspace*{9 mm}$Z_n(y)= \int\Pi_{i=1}^N dy_i \delta(y_n - y){\exp{[-\beta U(y^N)]}}$

\hspace*{15 mm}$Z_n(0)= \int\Pi_{i=1}^N dy_i \delta(y_n - 0){\exp{[-\beta U(y^N)]}}$

are the constrained partition function integrals, $\delta$ is the Dirac function
and $\beta=(k_BT)^{-1}$.

For the PBD model the calculation of a partition function integral reduces 
to multiplication of N matrices. The discretization of the co-ordinate
variable and introduction of a proper cut off on the maximum values of $y's$ 
determines the size of the matrices. We have taken $-2\rm\AA$ and $120\rm\AA$
as the lower and upper limit of integration for each co-ordinate variable
and discretized space using the Gaussian-Legendre method with number of grid
points equal to 900. Note that the values of the partition function
integrals and therefore the values of $V(y)$ are independent of the 
limit of integration. We show in Fig.1 the value of $ V(y)$ as a function
of $y$ at $45^{0}C$. At a separation $y$ the base pair feels a drift force 
$F= -{\partial V(y)}/{\partial y}$ towards the origin $y=0$. As shown in the
inset of Fig.1
this force has a minimum at $y=0.2\rm\AA$. This minimum corresponds to
a force barrier which has been observed in theoretical investigation of force
induced unzipping of a dsDNA in the constant extension ensemble [16,17,18] and is 
attributed to a combination of the force needed to break the hydrogen
bonds and the force needed to  overcome the entropic 
barrier of the stacking interaction [16].

The dynamics of the base pair may be described by the Langevin equation
\ba
\frac{ dy}{dt}=-\Gamma\frac{ dV(y)}{dy}+\xi\hspace*{3mm};
\hspace*{12mm} <\xi \xi>  (t)=2\Gamma k_{B}T \delta (t)
\ea

where $\Gamma$ is a transport coefficient of dimension time/mass and
$\Gamma k_BTa_{AT}^2$ of dimension 1/time. Eq.(3)
describes a one-dimensional random walk in a potential $V(y)$. We use 
$a_{AT}$ and $\Gamma k_{B}Ta_{AT}^2$ to make, respectively, distance 
and time dimensionless. The Fokker-Planck equation corresponding to (3)
is found to be
\ba
\frac{\partial P}{\partial t}=\frac{\partial}{\partial y}\left[\frac{-\partial 
\beta V(y)}{\partial y}P\right]+\frac{\partial^2 P}{\partial y^2}
\ea

where $P(y,y_0;t)$ is the probability density of the random walkers.

We assume that if separation $y$ reduces to zero at time $t^\prime$, it will
not contribute to autocorrelation function defined as $C(t)=<y(t)y(0)>-<y>^2$
for $t>t^\prime$ and similarly any new fluctuational opening which appear
after $t=0$ will not contribute to $C(t)$. Thus for purposes of computing
the autocorrelation function we place an absorbing wall at $y=0, {\it i.e.}
P(y=0,t)=0$. In addition to this we may require $P(y=L,t)=0$, where $L$
depends on the size of the dsDNA molecule or on any other condition which
limits the size of the bubble. The problem of calculating the 
autocorrelation function $C(t)$ therefore reduces to finding how many 
walkers of an ensemble of random walkers distributed according to 
thermal equilibrium distribution at $t=0$ are still present at time 
$t$ and have not been absorbed by the wall at $y=0$ [19].

When a substitution $P={\it exp}[-\beta V(y)/2]\psi$ is used 
Eq.(4) reduces to
\ba
-\frac{\partial \psi}{\partial t}=H\psi,
\hspace*{8mm} 
H=-\frac{\partial^2}{\partial y^2} + v(y)
\ea
where
\ba
v(y)=\frac{1}{4}\left[\frac{\partial \beta V(y)}{\partial y}\right]^2 - 
\frac{1}{2}\left[\frac{\partial^2 \beta V(y)}{\partial y^2}\right]
\ea

This is the imaginary time Schr\"{o}dinger equation for a particle
of mass {1/2}  in the potential $v(y)$. Let $\phi_m(y)$ denote
the eigenfunctions of the operator $H$, $H\phi_m=E_m\phi_m$, with
$\phi_m(y=0)=0$ and $\int dy\phi_m^\star (y)\phi_{m^\prime}(y)=\delta_{mm^\prime}$ 
Then expanding $\psi(y,t)$ in terms of eigenfunctions $\phi_m$ and
using the initial  condition $P(y,y_0,t=0)=\delta(y-y_0)$
the transition probability from initial separation $y_0$ to a 
final separation $y$ at time $t$ is found to be 
\ba
P(y,y_0,t)={\it exp}\left[-\beta \frac{V(y)+V(y_0)}{2}\right]
\sum_m e^{-E_mt}\phi_m(y)\phi_m^\star(y_0)
\ea

For initial distribution of separation $y_0$ we choose the Boltzmann factor 
$B(y)=A{\it exp}(-\beta V(y))$ where $A=1/\int_0^Ldy_0{\it exp}(-\beta V(y_0))$
is a normalization factor. If we start with the equilibrium distribution 
function at time $t=0$, the distribution function at time $t, P(y,t)$ is
\ba
P(y,t)=A\sum_m e^{-E_mt}\int_0^L dy_0 {\it exp}\left[-\beta\frac{V(y)+V(y_0)}{2}
\right]\phi_m(y)\phi_m^{\star}(y_0)
\ea
$P(y,t)$ measures the survival probability.

For the autocorrelation function we get 
\ba
C(t)=\int_0^L P(y,t) dy
=A\sum_m e^{-E_mt}{\left|\int_0^L e^{-\beta V(y)/2}\phi_m(y)dy\right|}^2
\ea

The values of $\phi_m(y)$ and $E_m$ of the operator $H$ in Eq.(5) are 
determined numerically using a method developed by Sethia et.al.[20]. As shown
in Fig.2(a), $v(y)$ is attractive at small $y$, rises to a (repulsive)
maximum at $y=0.2\rm\AA$ and then decays to zero as $y$ increases.
The maximum in $v(y)$ corresponds to the minimum in $F$ shown in Fig.1. For
small values of m , $\phi_m(y)$ remains confined (see Fig.2(b)) in a 
region of separation which values are  smaller than the length of the molecule L. After
the first three eigenvalues which values are $E_0=0.0028,E_1=0.0080,
E_2=0.0125,E_m$ is found to increase with $\Delta E_m=E_{m+1}-E_m\simeq 0.0043$
for $m\leq50$. The free particle in a box like behaviour is found only 
after $m>100$ and therefore the values of $\phi_m$ and $E_m$ depend
on the value of $L$ only after $m\geq100$.
We have varied L from $80\rm\AA-120\rm\AA$
and found that the values of $C(t)$ and $P(y,t)$ do not change. The values
given in Fig.3 and Fig.4 correspond to $L=100\rm\AA$ which approximately
measures the length of the dsDNA molecule of 29 base pair.

In Fig.3 the rescaled autocorrelation function $g(u)=G(t/t_{1/2})$
where $G(t)=C(t)/C(0)$ and $t_{1/2}$ is such that $G(t_{1/2})=0.5$
[$4$], is plotted as a function of rescaled time $t/ t_{1/2}$.
When this figure is compared with the one given in [$4$] we find
a very good agreement over the entire time range of measurement. If we choose
$\Gamma k_BTa_{AT}^2=10^5 s^{-1}$ and plot $G(t)$ as a function
of time $t$ in $\it ms$ the resulting curve is found to be
in very good agreement with the corresponding curve given in [$4$].

In Fig.4(a) we show the distribution function $P(y,t)$ which gives the 
probability of separation $y$ of the tagged base pair at time $t$.
From the figure we find that the most probable separation is 
$y^\star\sim 1\rm\AA$,
although the term ``most probable" makes less and less sense
because the peak gets broader and broader. Thus, initially as well as
presently small separation of the order of $1\rm\AA$ make the most contributions to the 
autocorrelation function $C(t)$ at all times. This can be 
understood from the nature of the drift force $F(y)$ (shown 
in Fig.1) which favours small separation. Since small separations
have larger Boltzmann weights initially, they dominate $C(t)$ at all times.
If we plot $\it{ln}y^\star$ {\it vs} $\it lnt$ we find a straight line having a slope equal to
1/6. Thus the most probable displacements of the base pair depends
on time as $y^{\star}\sim t^{1/6}$.

The half time history of a random walker that is at $y^\star$ at $t=0$
and at $t$ is defined as [19]
\ba
H(y,t/2|y^\star,t;y^\star,t=0)=P(y^\star,t|y,t/2)P(y,t/2;y^\star,t=0)
=\left|\sum_m e^{-E_mt}\phi_m^\star(y^\star)\phi_m(y)\right|^2
\ea
We plot the half time distribution as a function of $y$ for $t=5,10$ and  $20$
in Fig 4(b). While values of $y^\star$ corresponding to these times
are $0.84,1.01$ and $1.24$ the peak in $H$ are found respectively at 
$0.90,1.12$ and $1.48$ which are somewhat larger 
than the corresponding values of $y^\star$.The half width of the distribution
 $H(y,t/2|y^\star,t;y^\star,0)$ is found to be narrower than that of $P(y,t)$. 
Therefore the most probable way for a displacements 
of size $y^\star$ formed at $t=0$ to survive until a time $t$ is that they 
first grow larger than $y^\star$ and then shrink back to the original
size.

In conclusion; we developed a theory to describe the multi-state 
relaxation dynamics of a tagged base pair of dsDNA. We used 
a potential model which describes interactions in dsDNA
at base pairs level and calculated the drift force which acts on
the base pair and drives it to its equilibrium position.
The dynamics is governed by the Langevin equation with Gaussian
white noise. We derived the associated Fokker-Planck equation
and with suitable transformation reduced it to an imaginary 
time Schr\"{o}dinger equation for a particle of mass $1/2$.
We found the eigenvalues and eigenfunctions of the operator
using a numerical method described in [20]. The calculated 
displacement autocorrelation function is found to agree
with experimental result for the entire time range of measurement.
The most probable displacements which contribute predominately
to short as well as long times are found to be small, of the order of
$1\rm\AA$. The half time
distribution of these displacements which show how the most
probable displacements behave between time $t=0$ and $t$
are calculated. The method developed here is equally 
applicable to homogeneous and heterogeneous DNA molecules.

Acknowledgments: We thank Navin Singh for his help in computation
and A. K. Ganguly for useful discussions. This work is supported 
by a research grant from DST of Govt. of India, New Delhi.

\hspace*{1in}
%-------------------------------------------------------------------------

\hspace*{1in}
\begin{figure}[h]
\includegraphics[width=3.5in]{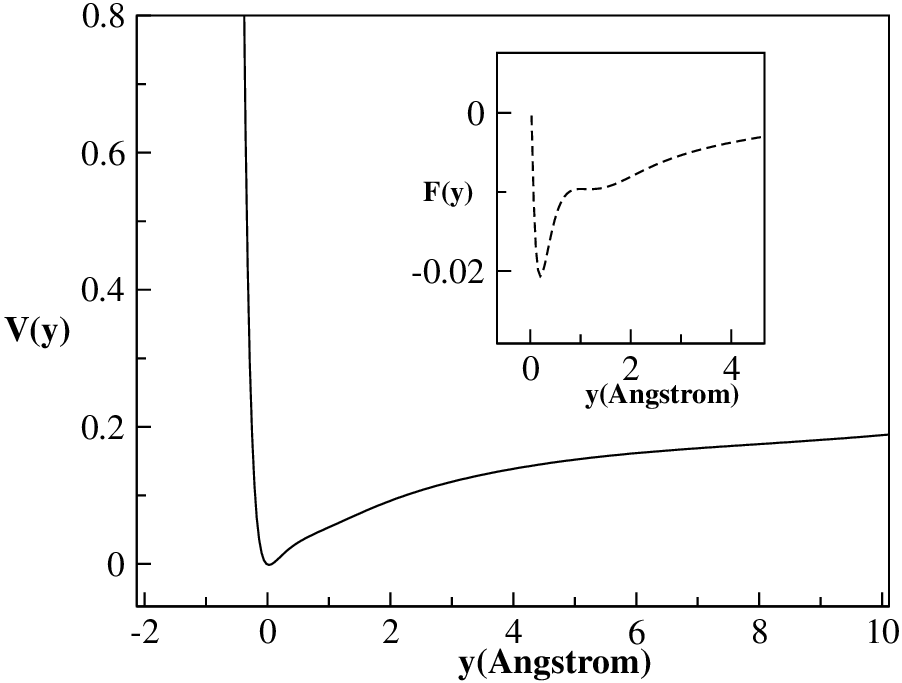}
\caption{The effective potential felt by the tagged base pair in a 
dsDNA molecule (A18 of [4]) of 29 base pairs at separation $y$ at $45^oC$.
In the inset the drift force $F(y)=-{\partial V(y)/\partial y}$   
which drives the base pair to the equilibrium position is plotted.} 
\end{figure}

\begin{figure}[h]
\includegraphics[width=3.5in]{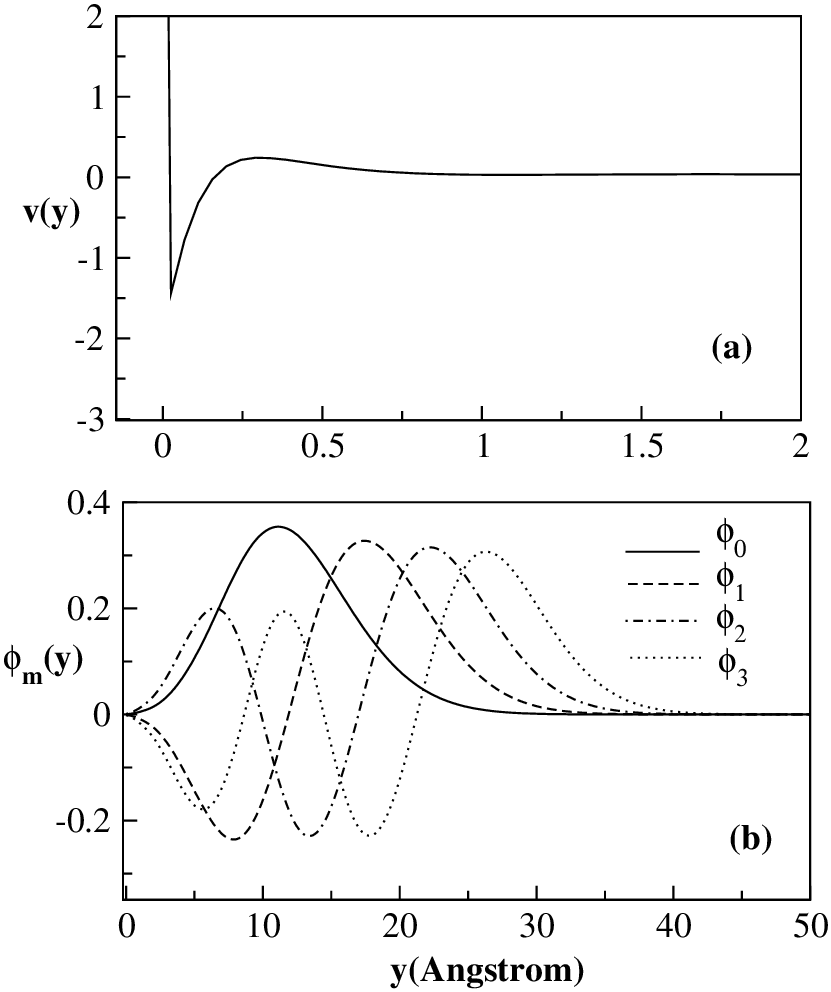}
\caption{{\bf (a)}The potential $v(y)$ of Eq.(6) at $45^oC$. The (repulsive)
maximum in $v(y)$ corresponds to the minimum in the drift force $F(y)$.
{\bf (b)} Results of first few eigenfunctions as a function of $y$. }
\end{figure}
\vspace*{1in}

\begin{figure}[h]
\includegraphics[width=3.5in]{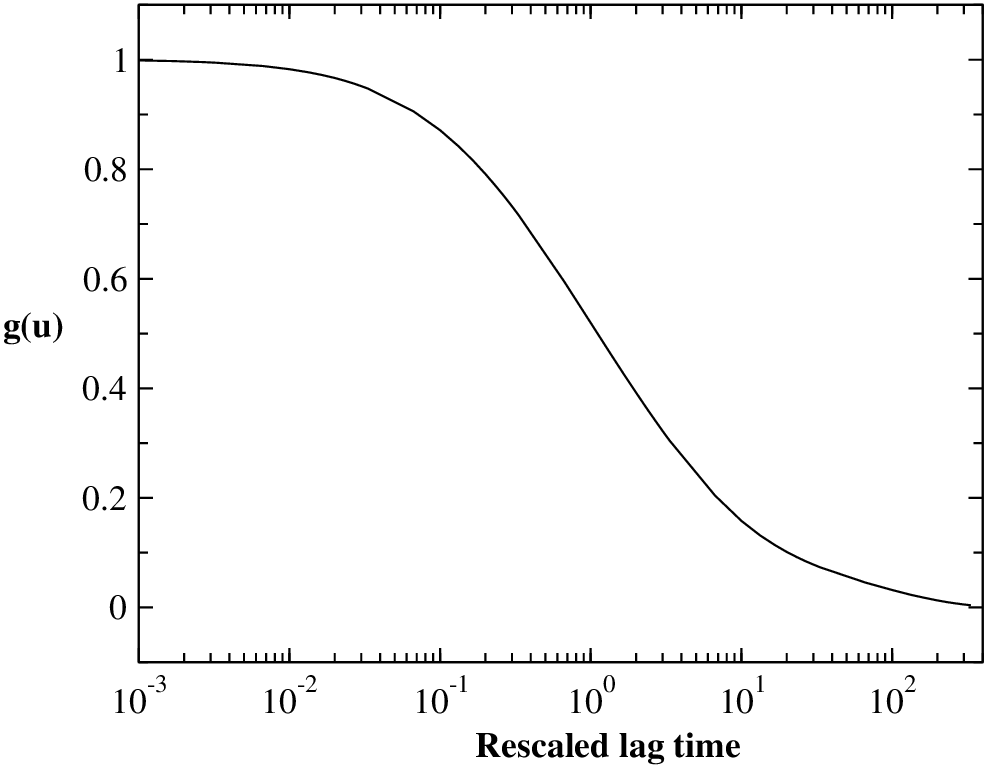}
\caption{Rescaled autocorrelation function $g(u)=G(t/t_{1/2})$ where 
$G(t)=C(t)/C(0)$ and $t_{1/2}$ is such that $G(t_{1/2})=0.5$ as a 
function of $t/t_{1/2}$ at $45^oC$.These notations are same as used in [4].}
\end{figure}

\begin{figure}[h]
\includegraphics[width=3.5in]{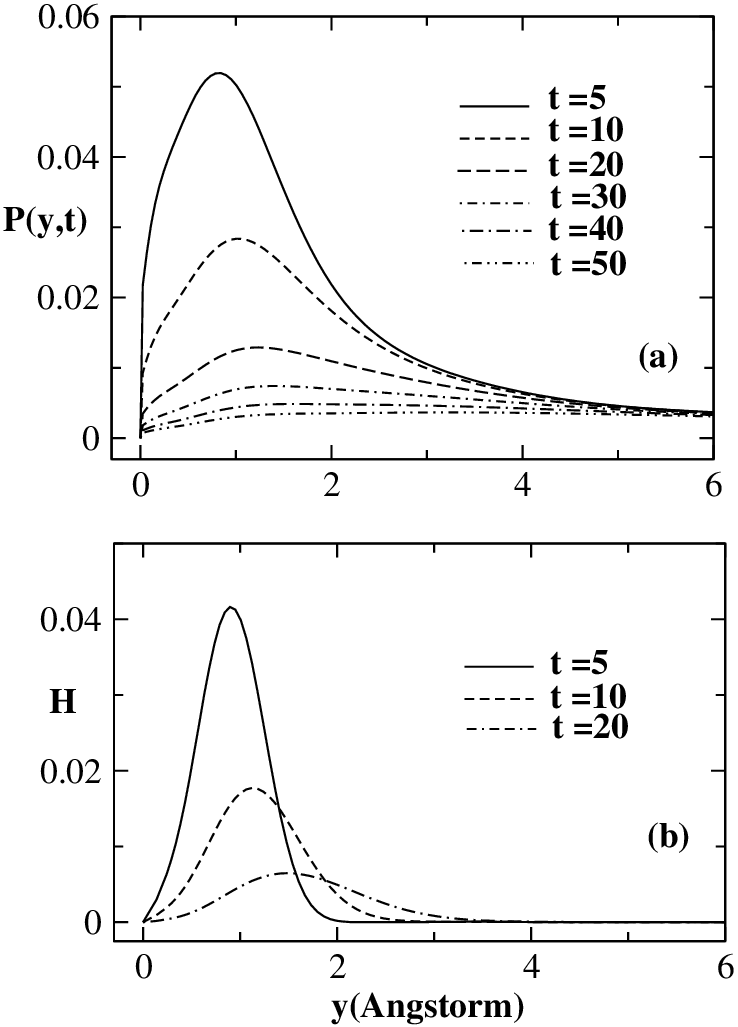}
\caption{{\bf (a)}Results for the distribution $P(y,t)$ as a function of
separation $y$ at $45^oC$ for several time $t$ which are expressed 
in unit of $(\Gamma k_BTa_{AT}^2)^{-1}$. The peak in $P(y,t)$ represents the
most probable separations. {\bf (b)} Results for the
half-time distribution $H(y,t/2|y^\star,t;y^\star,t=0)$ as a function of separation 
$y$ for $t=5,10,20$ for which $y^\star=0.84,1.01,1.24$. The peak in $H$
is found at separation larger than the corresponding value of $y^\star$.}
\end{figure}
%-------------------------------------------------------------------------

\begin{thebibliography}{99}
\bibitem{1}  W. Saenger, {\it Principle of Nucleic Acid Structure} (Springer Verlag,
Berlin,1984).
\bibitem{2} J. Sontalucia Jr., {\it Proc Nat. Acad. Sci., U.S.A. } {\bf 95}, 1460 (1998);
F.Pincet, E. Perez, G. Bryant {\it et al.}, Phys. Rev. Lett. {\bf 73}, 2780 (1994); 
A. Krueger, E. Protozanova and M. D. Frank-Kamenetskii, Biophys. J. {\bf 90}, 3091 (2006).
\bibitem{3} A. Campa, Phys. Rev. E {\bf 63}, 021901 (2001), M. Peyrard, 
Europhys. Lett. {\bf 44}, 271 (1998).
\bibitem{4} G. Altan-Bonnet, A. Libchaber and O. Krichevsky 
, Phys. Rev. Lett. {\bf 90}, 138101 (2003).
\bibitem{5} M. Gu\'{e}ron, M. Kochoyan and J. L. Leroy, {\it Nature} (London)
{\bf 328}, 89 (1987).
\bibitem{6} R. M. Wartall and A. S. Benight, Phys. Rep. {\bf 126}, 67 (1985).
\bibitem{7} A. Kornberg and T. A. Baker, {\it DNA Replication} (W. H. Freeman, NewYork, 1992).
\bibitem{8} R. J. Robert and X. Chang, Annu. Rev. Biochem. {\bf 67}, 181 (1998); 
J. T. Stivers, {\it Nucleic Acid Res Mol. Biol.} {\bf 77}, 37 (2004);
 J. F. L\'{e}ger {\it et al.}, {\it Proc Nat. Acad. Sci., U.S.A. }{\bf 95}, 12295(1998).
\bibitem{9} H. C. Fogedby and R. Metzler, Phys. Rev. Lett. {\bf 98}, 070601
(2007);Phys. Rev. E {\bf 76}, 061915 (2007).
\bibitem{10} D. Poland and H. A. Scheraga,{\it Theory of Helix-Coil Transition in 
Biopolymers} (Academic Press, New York,1970).
\bibitem{11} C. Vanderzande, {\it Lattice Models of Polymers} (Cambridge
University Press, Cambridge, 1998)
\bibitem{12} D. J. Bicout and E. Kats, Phys. Rev. E {\bf 70}, 010902(R) (2004).
\bibitem{13} T. Ambj\"{o}rnsson {\it et al.}, 
Phys. Rev. Lett. {\bf 97}, 128105 (2006); Biophys. J. {\bf 92}, 2674 (2007).
\bibitem{14} M. Peyrard and A. R. Bishop, Phys. Rev. Lett. {\bf 62}, 2755 (1989); 
T. Dauxois, M. Peyrard and A. R. Bishop , Phys. Rev. E {\bf 47}, 684 (1993).
\bibitem{15} A. Campa and A. Giansanti, Phys. Rev. E {\bf 58}, 3585 (1998).
\bibitem{16} S. Cocco, R. Monasson and J. F. Marko, {\it Proc. Natl. Acad.},
U.S.A. {\bf 98}, 8608 (2001); Phys. Rev. E {\bf 65}, 041907 (2002).
\bibitem{17} N. K. Voulgarakis {\it et al.},
 Phys. Rev. Lett. {\bf 96}, 248101 (2006).
\bibitem{18} N. Singh and Y. Singh, Eur. Phys. J. E. {\bf 17}, 7 (2005), 
{\bf 19}, 233 (2006).
\bibitem{19} C. Tang, H. Nakanishi and J. S. Langer, Phys. Rev. A {\bf 40}
,995 (1989).
\bibitem{20} A. Sethia, S. Sanyal and Y. Singh, J. Chem. Phys. {\bf 93}, 7268 (1990).

\end{thebibliography}
\end{document}